# Orbital origin of magnetic moment enhancement induced by charge density wave in kagome FeGe

Shulun Han[1,#], Linyang Li[2,#], Chi Sin Tang[3,#], Qi Wang[4,#], Lingfeng Zhang[1,*], Caozheng Diao[3], Mingwen Zhao[5], Shuo Sun[1], Lijun Tian[1], Mark B. H. Breese[3,6], Chuanbing Cai[1], Milorad V. Milošević[7], Yanpeng Qi[4,8,*], Andrew T. S. Wee[6,9,*], Xinmao Yin[1,*]

[1]Shanghai Key Laboratory of High Temperature Superconductors, Department of Physics, Shanghai University, Shanghai 200444, China

[2]School of Science, Hebei University of Technology, Tianjin 300401, China

[3]The Singapore Synchrotron Light Source (SSLS), National University of Singapore, Singapore 117603

[4]ShanghaiTech Laboratory for Topological Physics, School of Physical Science and Technology, ShanghaiTech University, Shanghai 201210, China

[5]School of Physics, Shandong University, Jinan 250100, China

[6]Department of Physics, National University of Singapore, Singapore 117551, Singapore

[7]Departement Fysica, Universiteit Antwerpen, Groenenborgerlaan 171, B-2020 Antwerpen, Belgium

[8]Shanghai Key Laboratory of High-resolution Electron Microscopy, ShanghaiTech University, Shanghai 201210, China

[9]Centre for Advanced 2D Materials and Graphene Research, National University of Singapore, Singapore 117546

**Abstract:** Interactions among various electronic states such as charge density wave (CDW), magnetism, and superconductivity are of high significance in strongly-correlated systems. While significant progress has been made in understanding the relationship between CDW and superconductivity, the interplay between CDW and magnetic order remains largely elusive. Kagome lattices, which intertwine nontrivial topology, charge order, and magnetism, offer an ideal platform for such studies. The kagome magnet FeGe, hosting the unique coupling between CDW and magnetism, has recently garnered considerable attention in that respect. Here we reveal the significant role of the orbital coupling effect during the CDW phase transition, highlighting the orbital origin of the magnetic moment enhancement in FeGe. Our X-ray absorption experiments and first-principles calculations illuminate the temperature-dependent behavior of Fe$3d$—Ge$4p$ orbital hybridization and corroborate its pivotal impact on the magnetic properties of FeGe. These findings introduce an orbital dimension to the correlation between charge and magnetic degrees-of-freedom, advancing our understanding of the intriguing quantum phases resulting from this interplay.

**Introduction**

The coexistence and competition among multiple quantum phases lie at the forefront of research in Condensed Matter Physics. Such form of complex interplay is characterized by critical phenomena which take place at almost concurrent temperature scales(*1*) and they are particularly prominent in strongly-correlated systems including cuprates(*2*), iron pnictides(*3*), and heavy fermion compounds(*4*). Of particular interest are intricate interactions between charge order, magnetism and superconductivity, where the charge density wave (CDW) phenomena often coexist with the onset of superconductivity and magnetic phases(*5-8*). Notable instances could be observed in transition-metal dichalcogenides (TMDs) and cuprates, where the competition between superconductivity and CDW order is prominent(*5, 9*). While in cases of $R_5Ir_4Si_{10}$ and $RNiC_2$ (where R denotes rare earth atoms), the relationship between CDW with antiferromagnetic ordering is deeply inextricable(*10, 11*). The quest to unravel the intricate mechanisms behind these phenomena remains a challenging and ongoing endeavour.

In that quest, understanding the interactions of CDWs with other electronic orders is of crucial significance. Kagome materials, with their unique structural and electronic properties(*12, 13*), offer a particularly promising platform for investigating these complex relationships, having been predicted to host a diverse range of quantum states including superconductivity, topological Dirac cones and anomalous magnetic orders(*14-16*). A prominent research case is the non-magnetic kagome class of metals, $AV_3Sb_5$ (where A = K, Cs, Rb)(*17*), which has entered the spotlight due to its time-reversal symmetry breaking(*18*), pair density wave (PDW)(*19*), and coexistence of CDW and superconductivity(*20*). Similarly, FeGe, the first reported kagome magnet to exhibit CDW after magnetic order initiation(*21*), has also garnered strong interest. By employing a series of techniques including ARPES(*22*), neutron scattering(*21*), X-ray scattering(*23*), spectroscopic measurements(*24-26*), and scanning tunneling microscopy (STM)(*27, 28*), investigative studies have revealed a $2 \times 2 \times 2$ superlattice and unique coupling between CDW and magnetism in FeGe, offering much needed insights into the intricate interplay between these quantum phases.

The shared characteristics observed in FeGe and $AV_3Sb_5$ suggest a potential universality of the unconventional CDW occurrence in the kagome lattice(*29-31*). The most intriguing phenomenon is the significant enhancement of magnetic moments observed in FeGe after the CDW phase transition(*21, 22*). This indicates an intimate coupling between spin, charge, and lattice degrees-of-freedom. Despite these highly motivating insights, the exact nature of the interplay between CDW and magnetic orders remained elusive to date, mandating further investigations into the strongly-coupled CDW systems. FeGe, with its unique properties, serves as an ideal platform to advance such studies. As kagome materials continue to enrich our understanding of various quantum phases and instabilities, the interaction between charge and magnetism is poised to establish itself as an exciting research frontier.

With latter motivation, in this article we demonstrate the temperature-dependent behavior of orbital coupling associated with the CDW phase transition in FeGe, through a comprehensive scientific analysis that integrates X-ray Absorption Spectroscopy (XAS) and Density Functional Theory (DFT) calculations. By tracking atomic movements and the orbital hybridization dynamics, we are able to gain insights to the changes in orbital information during the evolution of lattice distortion. Our findings suggest the direct influence of orbital hybridization on the Fe3*d* orbitals on the unconventional enhancement of magnetic moments following the onset of the CDW order (Fig. 1A). These results offer a valuable dimension to understanding the relationship between charge and magnetic degrees-of-freedom, and advance our comprehension of the deeply intertwined electronic orders in quantum materials.

**Results**
Single crystals of FeGe were synthesized via chemical vapor transport(*21*). In its pristine phase, the FeGe crystal exhibits a hexagonal structure with space group P6/mmm, which is isostructural to FeSn and CoSn(*32*). The crystal comprises a kagome lattice of Fe atoms with Ge1 situated at the hexagonal center, as shown in the inset of Fig. 1B. This arrangement results in a densely packed atomic structure within the kagome layer, where both Fe-Fe and Fe-Ge bonds exhibit similar lengths and strengths(*33*). These kagome layers and Ge2 honeycomb layers are stacked alternately along the c-axis. Powder X-ray diffraction (XRD) pattern of FeGe single crystals at room temperature is shown in Fig. 1B, proving the high crystalline quality of the sample.

To further confirm the electrical and magnetic quality of our FeGe single crystal, temperature-dependent DC resistivity and in-plane magnetic susceptibility characterization of the sample are measured. Fig. 1C exhibits a typical metallic behavior across the entire temperature range, consistent with previous reports(*21, 24*). As seen in the inset of Fig. 1C, by calculating the first derivative of the sample resistivity ($d\rho/dT$), a kink is observed at ~112 K. This indicates the onset of a CDW transition(*21*). The magnetic susceptibility χ of our sample also shows a decrease around 110 K (Fig. 1D) associated with the CDW phase transition. Furthermore, with spin canting taking place at ~ 70 K, it leads to a rapid rise in magnetization, producing a maximized χ at ~30 K.

Kagome magnet FeGe exhibits a series of quantum phases. Namely, an A-type antiferromagnetic (AFM) phase sets in below 410 K(*21, 34-36*), characterized by AFM order between the layers and ferromagnetic (FM) alignment within each plane (see Fig. 1A). A short-range CDW order appears at ~100 K(*21, 27*), which can be extended into a long-range order through post-growth annealing(*37-39*). The CDW phase transition is accompanied by the anomalous Hall effect and the emergence of topological edge(*21, 27*), similar to observations in the previously reported $AV_3Sb_5$ system(*40, 41*). Below ~60 K, spin canting leads to the formation of a c-axis double cone AFM structure(*35, 36*), with additional spin-flip phase transitions induced by applying an external magnetic field(*21*). The orbital degree-of-freedom of electrons in kagome materials has not been sufficiently explored, although one knows that orbital hybridization, as influenced by crystal structures, local symmetries, and spin coupling, is often the driving force behind the novel phenomena in correlated electron systems. While previous studies have reported the presence of iron-germanium hybridization in FeGe(*28, 33, 42*), the specific role it plays has not been examined in detail. Beyond the confines of the FeGe system, such orbital coupling effects have also been seen in similar systems, including hexagonal FeSn(*43*) and CoSn(*44*), in which they contribute significantly to their novel physical properties. Therefore, the main objective of our investigation was to garner deeper insights into the role that orbital hybridization plays across different phases and amidst the phase transition processes, through spectroscopic techniques. Specifically, we employed XAS, a powerful element-specific investigative tool, to probe the unoccupied states, electronic structures and orbital coupling of the system concerned.

Fig. 2A shows the XAS spectra for Fe3*d* orbital in FeGe, displaying multiplet peak structures different from those of bulk iron metal(*45*). The two broad parts observed respectively at about 711 and 723 eV stem from the spin-orbit-split Fe2*p* states, representing Fe $L_3$ and $L_2$ edges for $2p_{3/2}/p_{1/2} \rightarrow 3d$ electron transitions. In order to investigate the orbital coupling properties during the CDW phase transition, we conducted the temperature-dependent XAS measurements on FeGe in the range between 50 and 200 K. As shown in Fig. 2A, the spectral intensity displays significant temperature-dependent fluctuations, which combined with the results from first-principles studies as presented later, signify the impact of orbital hybridization between Fe and neighboring Ge atoms. Due to the ligand field effect, the main parts further split into four peaks labeled as A (709.6 eV), B (711.2 eV), C (722.8 eV), and D (724.3 eV), arranged in ascending order of energy. A shoulder feature around 715 eV is observed on the right of peak B through the peak fitting analysis, yet it exhibits almost no fluctuations. Consequently, we will exclude this feature from subsequent analysis. One can see that there are two critical temperatures in the color plot of Fig. 2B, $T_{CDW}$ and $T_{Canting}$, clearly observable on the $L_3$ and $L_2$ edges, while the temperature dependence at other energy positions exhibits minor changes. Specifically, the spectral intensity around ~710 eV displays an initial enhancement starting from 200 K, followed by a weakened intensity at 110 K and below acting as a turning point. Below 70 K, a more rapid decline in intensity is evident.

Although the signal from the $L_2$ edge is typically considerably weaker, and the $L_3$ edge of Fe3*d* orbitals provides clearer and more accurate information on the system's hybridization dynamics, it is still important to analyze the fluctuation trends of both edges. Subsequently, we obtained the intensity of the four peaks at different temperatures, thereby delving into the temperature-dependent behavior of orbital hybridization. Fig. 2C illustrates the surprising finding that the evolution of peaks A, B, C and D follows a similar trend. Particularly notable is the abrupt decrease in intensity observed after $T_{CDW}$, indicating a significant change in orbital coupling associated with the magnetism or CDW phase transition in kagome FeGe. This unique behavior of CDW as a first-order phase transition in XAS spectra enriches our understanding of its dynamic properties, consistent with neutron scattering(*21*), Raman(*25*) and polarized infrared spectroscopy(*26*). For further analysis, we divided the temperature range into three characteristic regions, as indicated in Fig. 2C.

In the temperature region I, orbital coupling effects intensify with decreasing temperature. This is possibly attributed to the pronounced correlation effects (*22, 46*). Interestingly, this is a contrasting trend with the previous study involving $CsV_3Sb_5$, where orbital hybridization effects weakened with decreasing temperature, primarily due to the lack of magnetic interactions affecting the orbital dynamics(*47*). In comparison to $AV_3Sb_5$, FeGe displays stronger electron-electron interactions and spin-orbit coupling effects. As temperature decreases, electron thermal vibration diminishes gradually with a heightened Coulomb repulsion to promote electron localization(*48*), thereby intensifying the coupling amongst electron spin, orbitals, and charges. Since the initial states and their multiplicity distributions corresponding to the $L_3$ and $L_2$ edges are different, their responses to these effects vary. Consequently, this leads to a slightly different trend in absorption intensity as temperature changes.

We place a greater emphasis on monitoring changes in orbital coupling after the CDW transition temperature, where the magnetic moments of Fe atoms increase significantly (as highlighted by the yellow arrows in Fig. 2C). According to previous computational predictions(*49*), during the CDW phase transition, 1/4 of the Ge1 atoms undergo movement along the c-axis, resulting in a significant dimerization (with a total displacement of ~0.65 Å). The dimerization process involves the mutual approach of Ge1b atoms located at the corners of adjacent layers within a unit cell (Fig. 3A). Meanwhile, the displacements of Fe and other Ge atoms are minimal, and the Ge2 atoms experience Kekulé distortion (Figs. 3B and 3C). The atoms between adjacent layers move in opposite directions. It is noteworthy that in FeGe lattice distortion involves mainly the out-of-plane displacement of Ge1 atoms rather than the Fe atoms, which constitute the kagome network. The magnetic energy saving and structural energy engage in competition until the energy minimum equilibrium is reached, ultimately resulting in a $2 \times 2 \times 2$ CDW structural order(*49*). These behaviors contrast with the CDW ground state observed in non-magnetic $AV_3Sb_5$ materials(*50*), suggesting an origin different from the conventional CDW mechanism.

While experimental results have confirmed the coupling between the magnetic and CDW orders, whereby magnetic moments are effective enhanced below $T_{CDW}$(*21, 22*), the precise nature of their interaction remains unclear. Hence, it is crucial to elucidate their interaction from the perspective of orbital coupling, where orbital behaviors are considered the intermediary to induce such transition processes. Detailed first-principles calculations are conducted to scrutinize the evolution of the density of states (DOS) of the constituent atomic orbitals. The combined analysis of the experimentally-derived XAS spectra in tandem with theoretical calculations clearly points to the significance of iron-germanium hybridization near the Fermi level and highlights the role it plays in facilitating the CDW phase transition process.

The electronic and magnetic properties of a system are largely determined by the filling of orbitals near the Fermi level. To understand the magnetic band structure of FeGe, the partial density of states (PDOS) distribution near the Fermi level for the calculated Fe and Ge orbitals is shown in Fig. 3D, predominantly influenced by the contributions from Fe atoms. Interactions driving the magnetic splitting within each FM layer give rise to energy bands for both spin-minority (spin down) and spin-majority (spin up) electronic bands(*22*). These bands are symmetrically distributed in PDOS with respect to the energy axis, indicating that the system is antiferromagnetically ordered. Furthermore, the Fe PDOS is dominated by the Fe3*d* orbital (Fig. 3E), while the Ge PDOS is governed by the Ge4*p* orbital (see the supplementary materials for more details).

Due to the relatively insignificant structural distortions of other atomic components, the lattice distortion process that takes place during the CDW process in FeGe can be regarded as a linear motion of the Ge atoms, neglecting the displacements of other atoms. Hence, the calculation tracks the evolution of the Fe3*d* and Ge4*p* orbitals PDOS with respect to the displacement of Ge1 atoms during the phase transition process. Along with the temperature-dependent XAS experimental findings, our focus will extend to the behavior of orbital states located above the Fermi level.

Figs. 4A and 4B illustrate the PDOS of the Fe3*d* and Ge4*p* orbitals in the intermediate stage between the pristine to the CDW state, with parameter d denoting the displacement of Ge atoms along the positive/negative direction of the *c*-axis. Here, d=0 represents the pristine state of FeGe. The PDOS variation describes how the orbital hybridization effects evolve with temperature. A distinct feature (labeled *α*) is observed at ~2 eV in the PDOS belonging to the Fe3*d* orbital. This corresponds to the main feature B of the XAS spectra as displayed in Fig. 2A. This relation between the PDOS data and that of the XAS spectra could be derived due to the similar energy differences between features A and B (Position B - Position

A ≈ 1.5 eV) (Fig. 2A), and that of the energy difference between the features seen in the PDOS data. The peak labeled σ, situated ~1.5 eV away from peak α, therefore corresponds to feature A in the XAS spectra. By observing the change in the PDOS, it is noteworthy that the Ge4*p* component in the PDOS calculations exhibits significant fluctuations due to the large dimerization of the Ge atoms.

The most notable variations in the DOS occur for peaks α and σ, where the temperature-dependent fluctuations of features A and B of the XAS spectra are primarily attributed to their contributions. Therefore, greater emphasis will be placed on studying the evolution of these two peaks. With the displacement of Ge atoms, Figs. 4C and 4D show the relative intensity of peaks α and σ for Fe3*d* and Ge4*p* orbitals of the PDOS have been suppressed, leading to a weakened orbital hybridization. This accounts for the significant decrease in spectral intensity within temperature region II after CDW phase transition has taken place for FeGe. It further confirms that the reduction in peaks intensity in the XAS spectra is directly attributed to orbital hybridization effect. We observe that the Ge-4*p* orbital PDOS demonstrates a faster decrease as d increases from 0 to 0.6 Å. This is particularly noticeable for peak σ (Fig. 4B). Beyond that when d is increased from 0.6 to 0.8 Å, the PDOS exhibits a slower variation. This variation in the rate of decrease agrees with the initial rapid decline followed by a slow decrease observed experimentally in temperature region II. This suggests that the dimerization of the Ge atoms is almost completed at ~0.6 Å, and the CDW structure distortion primarily occurs in temperature region II – above 70 K.

Additionally, the subsequent rapid attenuation of hybridization effects in temperature region III may be due to alterations in the net magnetic moment induced by the double cone AFM structure(*35, 36*) below ~70 K. The abrupt changes observed in XAS experiments on our sample indicate that spin canting occurs around 70 K, which is slightly higher than previously reported(*21*). However, the influence of spin canting on orbital information remains to be examined in greater detail. Compared to the pristine structure, this atomic motion along the *c*-axis induces the formation of longer Fe-Ge bonds within the kagome layer (Fig. 3B). This may also be one of the reasons for the weakening orbital hybridization after structure distortion. Following the CDW phase transition, the enhanced intensity of magnetic moments on Fe atoms correlates with the weakening orbital hybridization, similar to iron-based materials(*51-53*). This analogous relationship underscores the potential for FeGe to serve as a platform for the exploration of unconventional superconductivity.

**Discussion**
Orbital hybridization has been demonstrated to exert profound effects on the 3*d* orbitals of metals(*54-56*). The introduction of orbital hybridization not only changes the properties of metallic 3*d* orbitals at the microscopic level but also directly influences their occupation. In pristine FeGe, the shorter bonds between Fe atoms demonstrate increased covalent properties. The significant variation in Fe wave functions and intense interactions between Fe*d* orbitals enhance the occupancy of the down-spin *d* electrons(*57*), resulting in decreased local magnetic moments. Additionally, the substantial hybridization between in-plane Fe3*d* and Ge4*p* orbitals significantly curtails the magnetic moment. Our results show that reduced orbital hybridization is intrinsically linked to the enhancement of magnetic moment in FeGe. With this understanding, we propose two scenarios to elucidate this phenomenon following the appearance of CDW order, both proving that the local orbital environment plays an important role in determining the magnetic moment of iron atoms in this material.

Firstly, strong hybridization is likely to be accompanied by electron hopping(*51, 58*),

providing a pathway to transitions between the Fe3$d$ and Ge4$p$ orbitals (Fig. 1A). Given the appropriate symmetry of in-plane Fe3$d$ orbitals hybridized with neighboring Ge4$p$ orbitals, such transitions primarily occur within the $d_{x2-y2}$ and $d_{xy}$ orbitals. The spin-down electrons occupying the Ge4$p$ orbitals transition to the Fe3$d$ orbitals, increasing the fully occupied states of the Fe3$d$ orbitals. This results in a reduction of the magnetic moment in pristine structure. As the temperature falls below T$_{CDW}$, the hybridization effect gradually weakens due to dimerization effects where the impact of the electronic transitions on the magnetic properties is also reduced. As a result, additional 3$d$ orbitals from Fe atoms become less occupied, leading to an increase in their magnetic moments.

Second, spin polarization in the Fe-atom is driven by the exchange interaction between Fe atoms(*59*). With the hybridization between the Fe3$d$ and Ge4$p$ orbitals in pristine structure, there is an increase in itinerant electrons on the Fe site (i.e., the number of 3$d$ holes decreases)(*60*). Therefore, the Fe magnetic moment is limited as the delocalization of the Fe3$d$ states. After the onset of CDW phase transition, the alternating out-of-plane motion of atoms increases the bond length between the atoms within the kagome layer, weakening the hybridization interaction. The Fe3$d$ states become more localized due to the larger distance from other neighboring atoms, thus exhibiting stronger magnetic properties.

In summary, we systematically investigated the temperature-dependent orbital coupling effect, emphasizing the orbital origin of the complex interplay between magnetic and charge order in FeGe. By tracing the evolution of orbital DOS during the dimerization process of Ge atoms, we infer the unusual weakening of hybridization between Fe3$d$ and Ge4$p$ below the onset temperature of the CDW transition. This suggests a close relationship between the enhancement of magnetic moments induced by the CDW and the rapid attenuation of orbital hybridization.

A significant new insight gained is that the interaction between CDW and magnetism can be described by the impact of orbital hybridization on Fe3$d$ orbitals, which determines the magnetic properties of the material. It is suggested that the hybridization effect in FeGe induces electronic transitions between Fe3$d$ and Ge4$p$ orbitals. As hybridization weakens, there is an increase of the 3$d$ orbital incomplete occupied states, leading to stronger magnetic moments. Alternatively, the enhancement of magnetism after the CDW phase transition can also be explained by the strengthening of localized Fe atom $d$-states, resulting from alterations in the electronic structure and the attenuation of hybridization effects.

The introduction of orbital degree-of-freedom thus provides novel yet very important insights into the intricate coupling phenomena between CDW and magnetic order in kagome FeGe. While numerous challenging questions persist in the understanding of quantum interplay in FeGe and similar materials, our findings offer a fresh perspective on the intertwined nature of magnetic and charge orders, opening new avenues for discussing the orbital origins and control of unconventional orders in solid-state systems.

FeGe. arXiv:2401.13474 (2024).

Fig. 1.

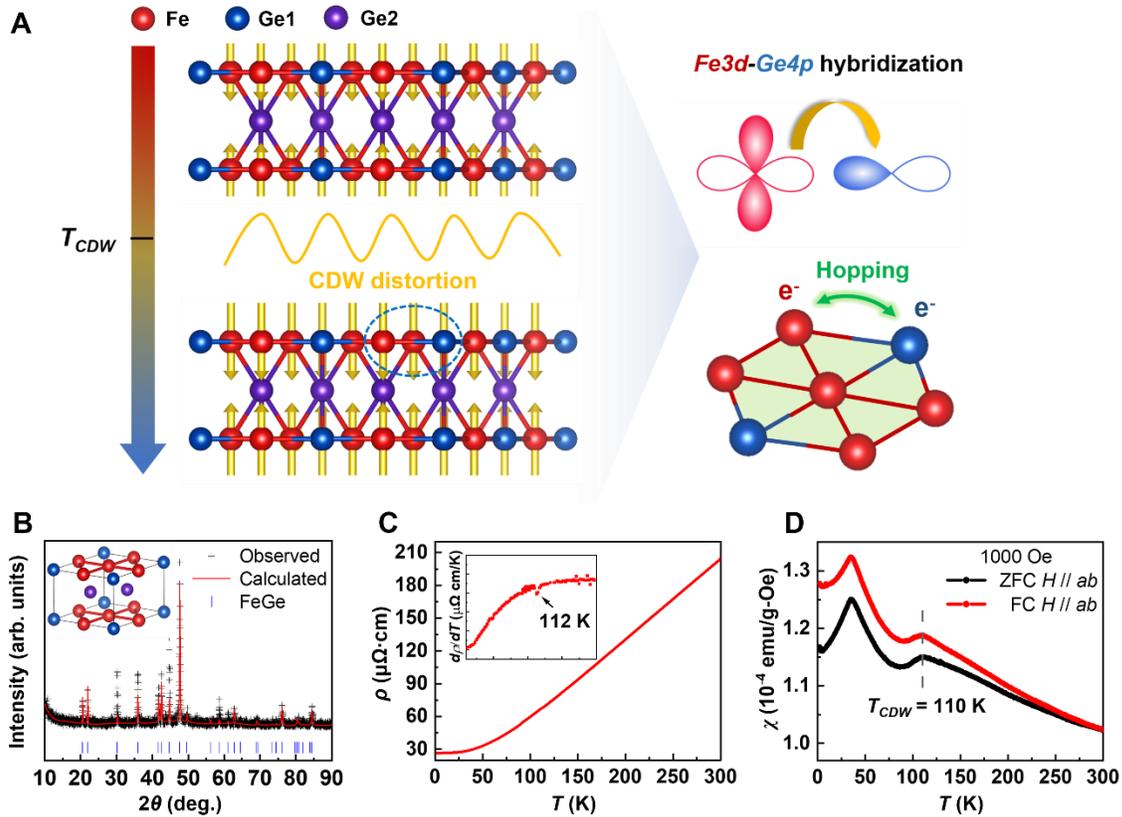

**Fig. 1. Novel phenomena accompanying the CDW transition and structural characterization of the FeGe single crystal samples.** (**A**) Schematic depiction of the changes in Fe3*d*—Ge4*p* orbital hybridization and magnetic moments during the CDW transition of FeGe. (**B**) Powder X-ray Diffraction (XRD) pattern. The calculated and experimental results show excellent consistency. Inset: unit pristine structure of FeGe. (**C**) Measured temperature-dependent DC resistivity of FeGe sample. Inset: the first derivative of the sample resistivity. (**D**) Magnetic susceptibility with ZFC and FC modes at $\mu_0 H$ = 1000 Oe for *H//ab*, results indicate the onset of CDW transition at ~110 K.

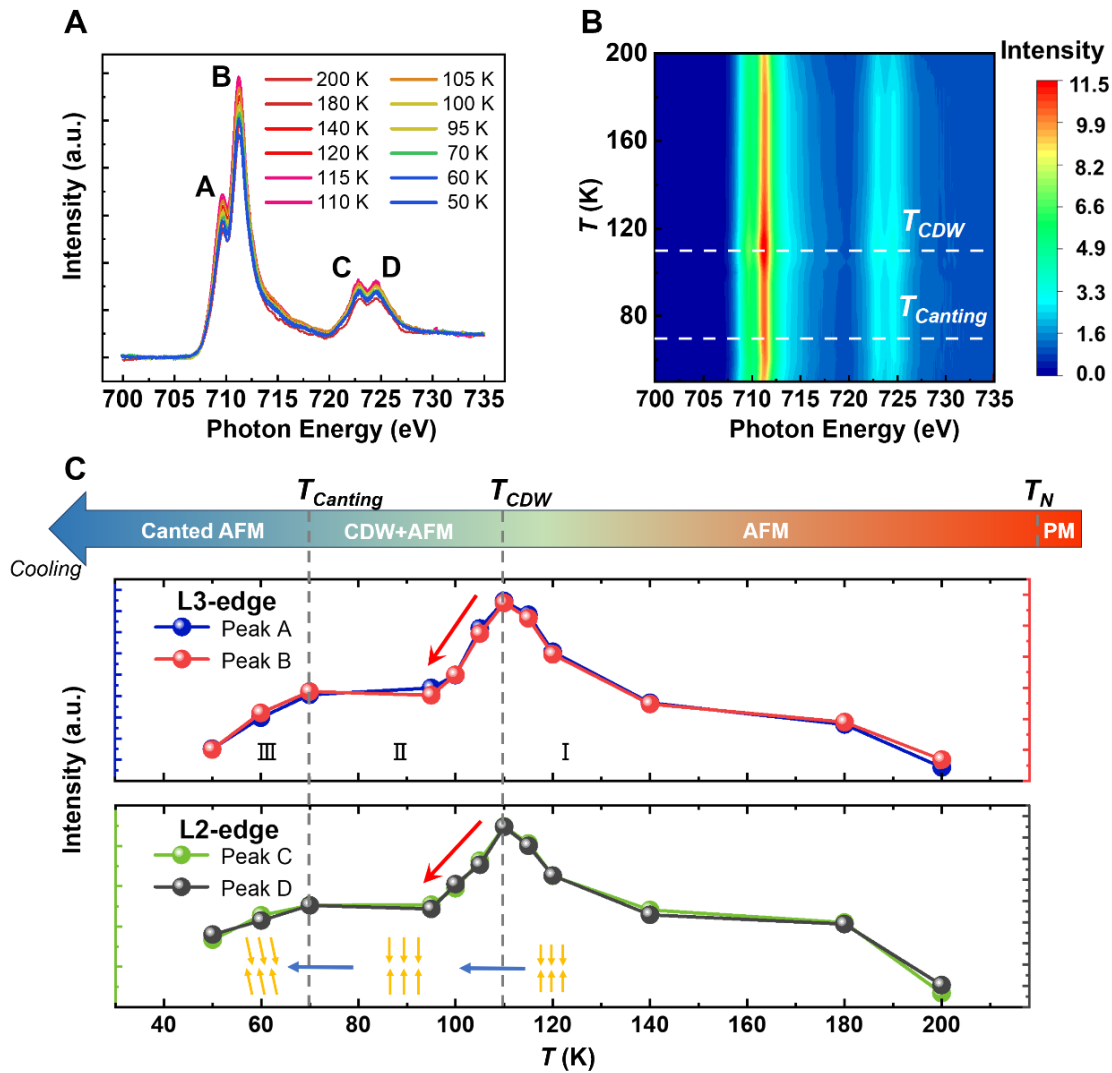

**Fig. 2. Temperature-dependent XAS characterization. (A)** Temperature-dependent Fe $L_{2,3}$-edge XAS spectra. **(B)** Color plot of Temperature-Energy phase diagrams by using the XAS data shown in (A). **(C)** Temperature-dependent intensity of features A, B, C and D across the three characteristic temperature regions I, II and III between 200 K and 50 K. The phase transition temperature axis of FeGe is shown above. Yellow arrows in the inset represent the changes in the magnetic moments of Fe atoms in each temperature region.

Fig. 3.

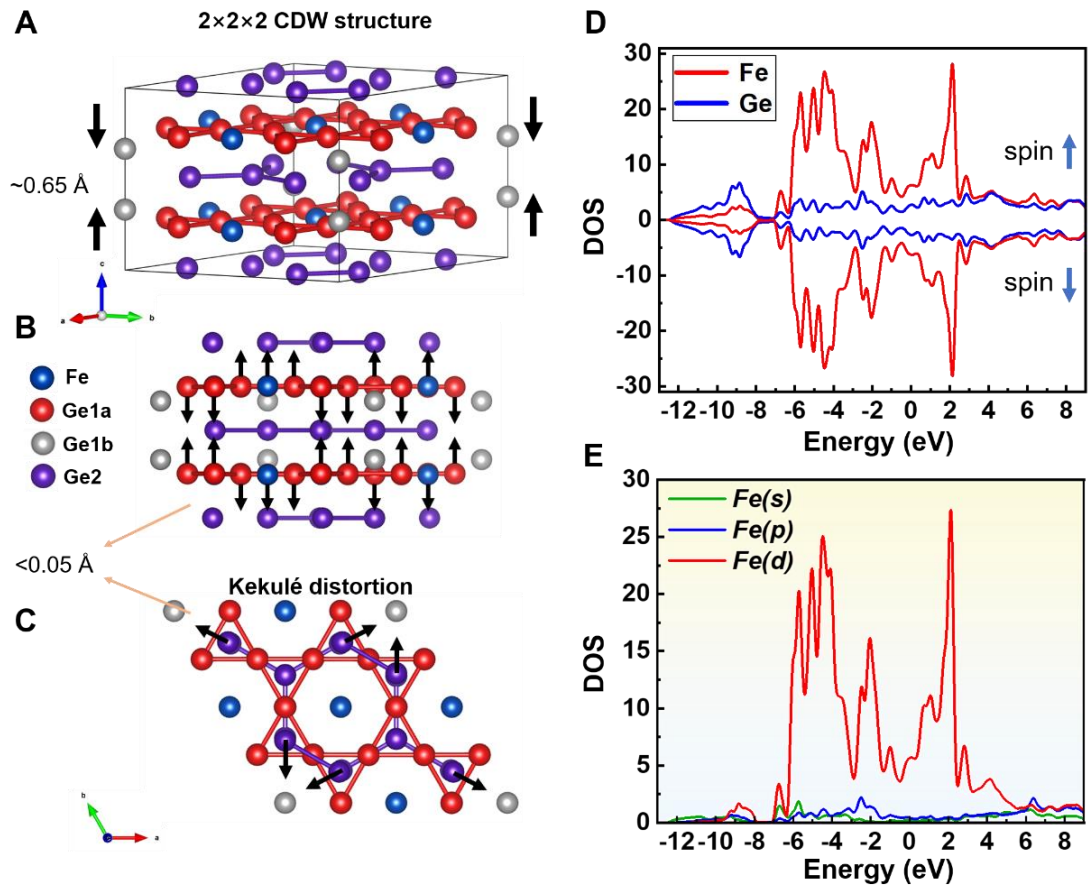

**Fig. 3. The calculated CDW structure and Density of States (DOS) of FeGe.** **(A)** 2 × 2 × 2 CDW superstructure with large 1/4 Ge1-dimerization, as indicated by the black longer arrows. **(B)** A side view of the CDW structure of FeGe. Black arrows are shown on the atoms to highlight the out-of-plane displacement from the kagome layers. **(C)** A top-down view of the CDW structure of FeGe. Black arrows are shown on the atoms to highlight the in-plane displacement within honeycomb planes. **(D)** Partial density of states (PDOS) calculated near the $E_F$ of the respective constituent orbitals of FeGe. **(E)** PDOS of $s$, $p$ and $d$ states of Fe in FeGe, where the $d$ orbital yields the dominant contribution.

Fig. 4.

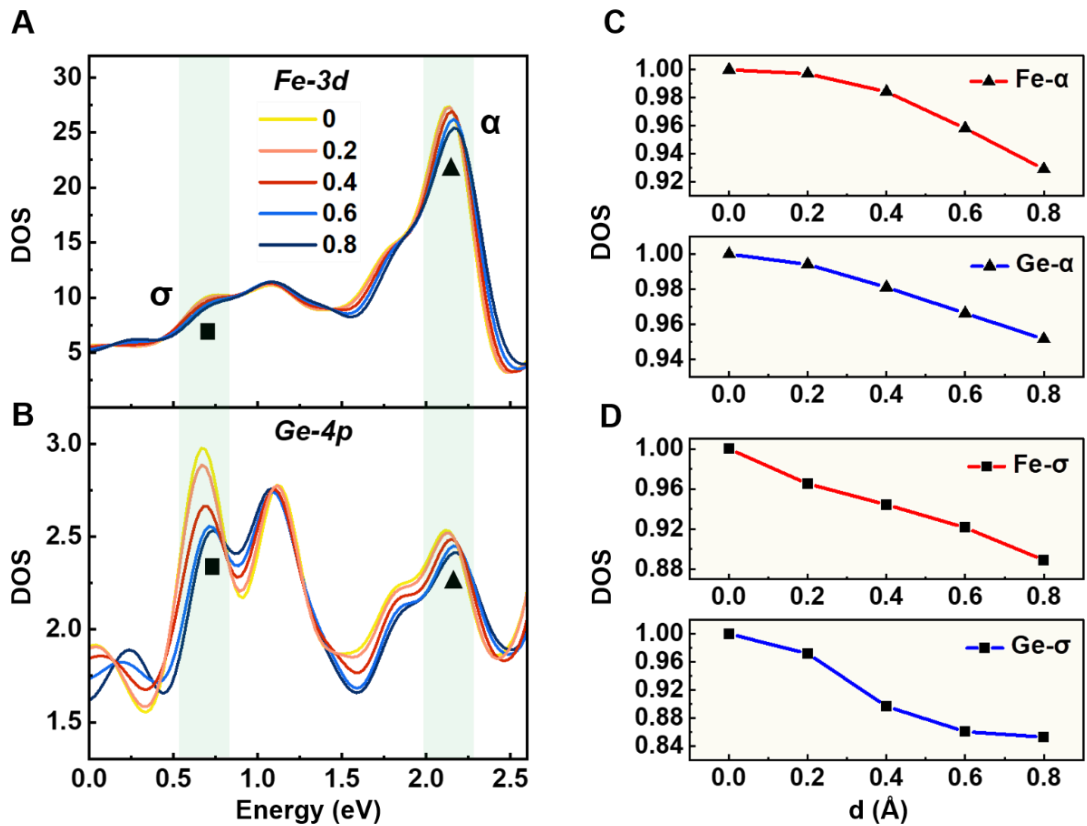

**Fig. 4. PDOS of the respective constituent orbitals for different lattice structures of FeGe.** **(A)** Changes in the PDOS of the constituent Fe$3d$, and **(B)** Ge$4p$ orbitals, as functions of the displacement of Ge1 (labelled d), during the intermediate CDW phase transition. Green shaded areas highlight the α and σ peaks, which make prominent contributions. Panels **(C)** and **(D)** show the changes in the relative intensities of the α and σ peaks, respectively, of the PDOS for both Fe$3d$ and Ge$4p$ orbitals during the intermediate CDW phase transition.